# Spectral comparison between AGN at $z$ = 0.1, 0.2 and 0.3

**Hartmut Winkler and Jessica Tsuen**

Department of Physics, University of Johannesburg, South Africa

E-mail: hwinkler@uj.ac.za

**Abstract**. We identified three samples of ROSAT sources with Sloan Digital Sky Survey spectra, one at redshift $z$ = 0.1, a second one at $z$ = 0.2 and a third one at $z$ = 0.3. 812 sources in total were examined. We determined the nature and spectral sub-types of the sources by visual inspection. The fraction of each sub-type at each of the three redshifts are then calculated. We consider selection biases caused by the luminosity cut-off threshold to determine whether any systematic trends in AGN type are evident with increasing redshift. We hence probe if an evolution effect is detected in our sample.

## 1. Introduction

The list of confirmed active galactic nuclei (AGN) has been dramatically boosted by a series of recent large scale spectroscopic surveys. In particular, the Sloan Digital Sky Survey (SDSS) has produced a vast number of optical spectra of extragalaxtic sources [1]. The resultant expanding database enables more detailed and rigorous study of the density and distribution of AGN in at least the comparatively nearby universe.

Despite the popularity and wide acceptance of the AGN unified model, which postulates that a common physical mechanism underlies all AGN types and that apparent differences can usually be ascribed to orientation effects, there is now also recognition that some AGN subclasses manifest real rather than apparent differences compared to other AGN. This then raises the question of whether the relative concentration of the different subclasses is a function of evolution, and hence effectively of distance, which is measured in terms of the redshift $z$.

Another aspect that becomes important in comparative studies of AGN subclasses is the relationship between their luminosities. While the range in AGN luminosities is vast, encompassing everything from the micro-AGN believed to exist in many otherwise ordinary galaxies to the most energetic distant quasars, some subclasses do on average appear to be intrinsically brighter than others.

In this work we attempt such a comparison of AGN subclass luminosities using a simple approach based entirely on visual spectral classification and population counts. We analyse spectra for sets of AGN identified in three 'shells' corresponding to redshifts of approximately $z$ = 0.1, 0.2 and 0.3. In order to circumvent selection biases caused by optical colour or spectral line effects, we defined our samples on the basis of x-ray detection. We recognise that x-ray detectability is in itself strongly dependent on redshift, but the x-ray luminosity bias should influence all subclasses equally, and thus not affect the ratio between subclass object counts.

## 2. Sample selection and classification of spectra

We started by making a positional cross-correlation of the x-ray sources detected in the ROSAT survey [2] with SDSS objects with spectra marking redshifts in one of the following three redshift ranges: $0.10 < z < 0.11$, $0.19 < z < 0.20$ and $0.29 < z < 0.30$. This resulted in a sample of 812 extragalactic objects, with at least 200 in each of the three ranges. Note that at greater z there is more contamination from the host galaxy spectrum, but this will not affect the AGN classification.

Spectra presented on the SDSS Data Release 10 website [1] were inspected and classified according to the scheme utilised in the ZORROASTER catalogue developed by one of us [3]. We point out that such visual inspections are able to distinguish between subtle spectral differences that are easily missed by automated line fitting and emission flux searches. A total of 379 of the initial 812 objects (= 47%) were thus confirmed to be AGN with Seyfert-like spectra. The remaining spectra either only exhibited weak starburst or weak LINER-like emission, or displayed no emission lines at all (meaning that the x-ray source was probably erroneously linked to them).

Having classified the AGN, we grouped them where appropriate into one of six sets listed in Table 1 below. Objects that could not be clearly assigned to one of these sets were not considered further.

**Table 1**. Distribution of AGN sub-types of ROSAT-detected galaxies in the three redshift ranges investigated.

| Object class | $z =$ 0.10-0.11 | $z =$ 0.19-0.20 | $z =$ 0.29-0.30 |
|---|---|---|---|
| Total ROSAT sources | 338 | 269 | 205 |
| Total ROSAT AGN | 115 | 142 | 122 |
| standard Seyfert 1 | 41 | 48 | 44 |
| NLS1 (weak Fe II) | 11 | 14 | 5 |
| I Zw 1 type objects | 8 | 17 | 18 |
| strong Fe II spectrum | 18 | 26 | 31 |
| peculiar Balmer profile | 2 | 9 | 4 |
| very wide Balmer lines | 3 | 9 | 9 |

The six sets may be described as follows. Two sets correspond to the so-called "narrow line Seyfert 1" subclass, or NLS1 [4]. We distinguish between such objects exhibiting a rich Fe II spectrum (often referred to as I Zw 1 objects after their prototype) and those NLS1 that do not have a similar iron line spectrum. Another set includes Seyferts with strong Fe II lines (but who are not NLS1). We then introduce sets for unusually wide Balmer lines and for Balmer lines with complex, asymmetric profiles often including red and/or blue bumps [5]. The final set is for 'ordinary' Seyfert 1 galaxies, i.e. objects with no distinct Fe II, He II or coronal line features, unremarkable Balmer broad line profiles (neither very wide nor very thin, reasonably symmetric with no bumps), and no hybrid spectrum typical of starburst or LINER galaxies.

## 3. Analysis of AGN spectral type distribution

In order to determine whether the fraction of specific AGN subtypes varies between the three samples, we employed a chi-squared statistical test. In each case, we compared the set tested with the set for the 'ordinary' Seyferts.

The chi-squared test yields a value that signifies that probability that the fraction of AGN belonging to a set in one of the three redshift ranges is significantly different to the fraction for that set in another redshift range. The values obtained are listed as percentages in Table 2.

**Table 2**. Results of the chi-squared analysis of various AGN subtypes versus the ordinary Seyfert 1 samples. Values represent the probability that the samples are significantly different. Values above 90% are given bold.

|  | NLS1 (weak FeII) | I Zw 1 type objects | strong Fe II spectrum | peculiar Balmer profile | very wide Balmer lines |
|---|---|---|---|---|---|
| $z = 0.1$ vs $z = 0.2$ | 16% | 84% | 51% | **94%** | 86% |
| $z = 0.2$ vs $z = 0.3$ | **94%** | 33% | 68% | 78% | 15% |
| $z = 0.1$ vs $z = 0.3$ | **90%** | **93%** | 89% | 53% | 89% |

None of the values exceeds the commonly employed 95% significance test, but several comparisons satisfy the less rigorous 90% significance mark. In particular, it is 90-95% likely that (i) AGN with peculiar Balmer profiles are less common at $z = 0.1$, (ii) NLS1 with weak Fe II are less common at $z = 0.3$ and (iii) I Zw 1 objects are less common at $z = 0.1$. There is also a hint that AGN with wide Balmer lines or strong Fe II occur more often at greater $z$, but this cannot be fully confirmed. Alternatively, the results may highlight intrinsic differences in the average luminosities of several of the sets. Resolving this requires an independent investigation. The present study also did not consider galaxy environments and clustering properties

**Appendix. Description of interesting AGN identified**
In Figure 1 we display four of the more noteworthy spectra in the investigated sample, which are briefly described in the following subsections.

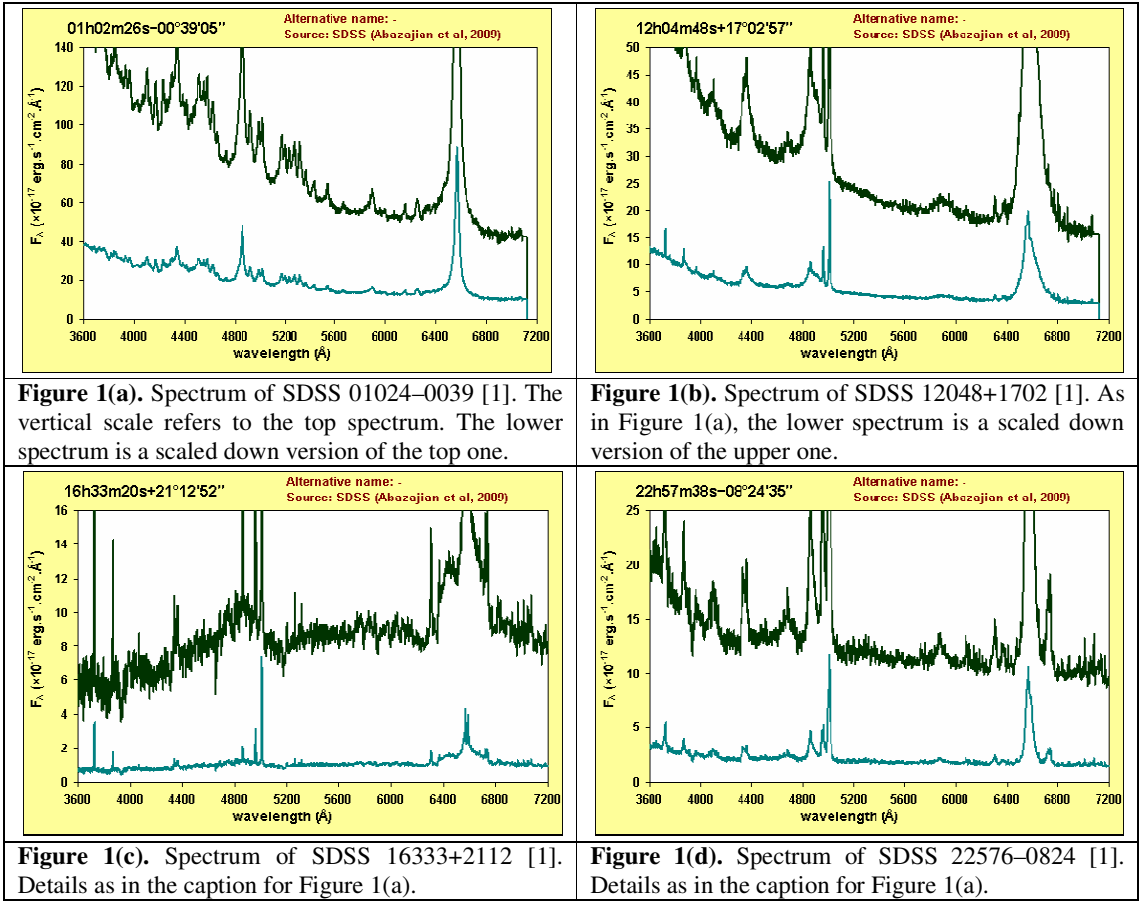

**Figure 1(a).** Spectrum of SDSS 01024–0039 [1]. The vertical scale refers to the top spectrum. The lower spectrum is a scaled down version of the top one.

**Figure 1(b).** Spectrum of SDSS 12048+1702 [1]. As in Figure 1(a), the lower spectrum is a scaled down version of the upper one.

**Figure 1(c).** Spectrum of SDSS 16333+2112 [1]. Details as in the caption for Figure 1(a).

**Figure 1(d).** Spectrum of SDSS 22576–0824 [1]. Details as in the caption for Figure 1(a).

3.1. SDSS 01024–0039
This is a well known nearby quasar, whose spectrum has been investigated in the past [6,7]. We note a very strong Fe II spectrum. No narrow lines are visible in Figure 1. The emission in the wavelength range in which the [O III] nebular lines are normally found are here in fact due to Fe II.

The spectrum in some degree resembles that of the nearby quasar Mkn 231 and the more recently discovered SDSS 12001–0204 [8], who also exhibit an exceptionally strong Fe II spectrum. Unlike these two objects, it does however not display a strong Na I absorption feature at 5892 Å.

### 3.2. SDSS 12048+1702

Figure 1(b) displays a Seyfert spectrum with very wide Balmer lines, which are in addition slightly asymmetric. In addition, the spectrum contains no detectable Fe II emission.

In contrast to other AGN featured in the ZORROASTER catalogue with a "w" descriptor for unusually wide lines, this object shows a distinctive He II 4686 Å broad emission feature.

### 3.3. SDSS 16333+2112

The spectrum of this object displays a complex Balmer line profile that is particularly striking in H-alpha (6562 Å). We note an H-alpha 'blue bump' clearly visible near 6400 Å, which would also be present in the remaining lines of the Balmer series. As the remaining Balmer lines are much weaker (both intrinsically, and also due to probable reddening), and due to greater contamination due to integrated starlight from the host galaxy, the 'blue bump' cannot be seen elsewhere.

The unusual profile is understood to be the result of the rotational dynamics of the AGN accretion disk and reverberation effects. This has been noted in a significant number of other AGN [5].

### 3.4. SDSS 22576–0824

This AGN exhibits comparatively narrow Balmer lines, only marginally wider than the forbidden lines. It is in that sense a typical example of a narrow line Seyfert 1 (NLS1), in this case without detectable Fe II emission.

What makes this object unusual is the strong He II 4686 Å emission, which requires exposure of the line emission region to abnormally powerful ultraviolet irradiance. The presence of this line is confirmed in some Seyfert 1's (those characterised by the descriptor "x" in the ZORROASTER catalogue [3]), but in a NLS1 this feature is rare.

**Acknowledgments –** This paper utilized data from the Sloan Digital Sky Survey (SDSS). Funding for the SDSS and SDSS-II has been provided by the Alfred P. Sloan Foundation, the Participating Institutions, the National Science Foundation, the U.S. Department of Energy, the National Aeronautics and Space Administration, the Japanese Monbukagakusho, the Max Planck Society, and the Higher Education Funding Council for England. The SDSS Web Site is http://www.sdss.org/.

The SDSS is managed by the Astrophysical Research Consortium for the Participating Institutions. The Participating Institutions are the American Museum of Natural History, Astrophysical Institute Potsdam, University of Basel, University of Cambridge, Case Western Reserve University, University of Chicago, Drexel University, Fermilab, the Institute for Advanced Study, the Japan Participation Group, Johns Hopkins University, the Joint Institute for Nuclear Astrophysics, the Kavli Institute for Particle Astrophysics and Cosmology, the Korean Scientist Group, the Chinese Academy of Sciences (LAMOST), Los Alamos National Laboratory, the Max-Planck-Institute for Astronomy (MPIA), the Max-Planck-Institute for Astrophysics (MPA), New Mexico State University, Ohio State University, University of Pittsburgh, University of Portsmouth, Princeton University, the United States Naval Observatory, and the University of Washington.